\documentclass[12pt,oneside,onecolumn]{IEEEtran}
\usepackage{amsmath,amsfonts,amssymb,amsbsy, amsthm,ae,aecompl}
\usepackage{algorithm,algpseudocode}
\usepackage[utf8]{inputenc}
\usepackage[english]{babel}
\usepackage{bm}
\usepackage{color}
\usepackage[noadjust]{cite}
\usepackage{epsfig}
\usepackage{enumerate}
\usepackage{float} 
\usepackage{fancyhdr}
\usepackage[T1]{fontenc}
\usepackage[acronym,toc,shortcuts]{glossaries}
\usepackage{graphicx, caption, subcaption}
\usepackage{hyperref}
\usepackage{lastpage}
\usepackage{listings}
\usepackage{lipsum}
\usepackage{dsfont}
\usepackage{multind}
\usepackage[normalem]{ulem}
\usepackage{transparent}
\usepackage{tikz,pgfplots}
\usepackage{times}
\usepackage{verbatim}
\usepackage[all]{xy}

\usetikzlibrary{calc}

\hypersetup{
    colorlinks,%
    citecolor=black,%
    filecolor=black,%
    linkcolor=black,%
    urlcolor=black
}

\pgfplotsset{
  grid style = {
    dash pattern = on 0.025mm off 0.95mm on 0.025mm off 0mm, % start with half a dot to get correct centering of the pattern
    line cap = round,
    black,
    line width = 0.5pt
  },
  tick label style={font=\small},
  label style={font=\small},
  legend style={font=\footnotesize},
}

\pgfplotscreateplotcyclelist{thechairLivingOnTheEdge}{
cyan!60!black,		densely dotted, 	every mark/.append style={fill=cyan!80!black},mark=otimes*\\%
red,                 solid, every mark/.append style={fill=red!80!black},mark=otimes*\\%
cyan!60!black, 		densely dotted,	every mark/.append style={fill=cyan!80!black},mark=diamond*\\%
red,           		solid, every mark/.append style={solid,fill=red!80!black},mark=diamond*\\%
teal,				every mark/.append style={fill=teal!80!black},mark=*\\%
orange,				every mark/.append style={fill=orange!80!black},mark=square*\\%
red!70!white,		mark=star\\%
yellow!60!black,		densely dashed, every mark/.append style={solid,fill=yellow!80!black},mark=square*\\%
black,				every mark/.append style={solid,fill=gray},mark=otimes*\\%
blue,				densely dashed,mark=star,every mark/.append style=solid\\%
red,					densely dashed,every mark/.append style={solid,fill=red!80!black},mark=diamond*\\%
}

\graphicspath{{figures/}}

\makeglossaries
\newacronym{CDN}{CDN}{content delivery network}
\newacronym{CF}{CF}{collaborative filtering}
\newacronym{D2D}{D2D}{device-to-device}
\newacronym{ICIC}{ICIC}{inter-cell interference coordination}
\newacronym{LTE}{LTE}{long term evolution}
\newacronym{SBS}{SBS}{small base station}
\newacronym{SCN}{SCN}{small cell network}
\newacronym{SVD}{SVD}{singular value decomposition}
\newacronym{QoS}{QoS}{quality-of-service}
\newacronym{QoE}{QoE}{quality-of-experience}
\newacronym{CRP}{CRP}{{C}hinese restaurant process}
\newacronym{RAN}{RAN}{radio access network}

\begin{document}
\title{Living on the Edge: The Role of Proactive Caching in 5G Wireless Networks}
\author{
		\IEEEauthorblockN{ Ejder Baştuğ$^{\diamond}$, Mehdi Bennis$^{\star}$ and Mérouane Debbah$^{\diamond}$,}\\
		\IEEEauthorblockA{
				$^{\diamond}$Alcatel-Lucent Chair - SUP\'ELEC, Gif-sur-Yvette, France \\	
				$^{\star}$Centre for Wireless Communications, University of Oulu, Finland \\
				\{ejder.bastug, merouane.debbah\}@supelec.fr, bennis@ee.oulu.fi
		}
		\thanks{This research has been supported by the ERC Starting Grant 305123 MORE (Advanced Mathematical Tools for Complex Network Engineering), the SHARING project under the Finland grant 128010 and the project BESTCOM.}
}
\maketitle
%\vspace{-100ex}
%\IEEEpeerreviewmaketitle

%%%%%%%%%%%%%%%%%%%%%%%%%%%%%%%%%%%%%%%%%%%%%%%%%%%%%%%%%%%%%%%%%%%
%%%%%%%%%%%%%%%%%%%%%%%%%%%%%%%%%% ABSTRACT
\begin{abstract}
This article explores one of the key enablers of beyond $4$G  wireless networks  leveraging small cell network deployments, namely \emph{proactive caching}. Endowed with predictive capabilities and harnessing recent developments in storage, context-awareness and social networks, \emph{peak} traffic demands can be substantially reduced by proactively serving predictable user demands, via caching at  base stations and users'  devices. In order to show the effectiveness of proactive caching, we examine two  case studies which exploit the spatial and social structure of the network, where proactive caching plays a crucial role. Firstly, in order to alleviate  backhaul congestion, we propose a mechanism whereby files are proactively cached during off-peak demands based on file popularity and correlations among users and files patterns. Secondly, leveraging social networks and device-to-device (D2D) communications, we propose a procedure that exploits the social structure of the network by predicting the set of influential users to (proactively) cache strategic contents and disseminate them to their social ties via  D2D communications. Exploiting this proactive caching paradigm, numerical results  show that important gains can be  obtained for each case study, with backhaul savings and a higher ratio of satisfied users of up to $22\%$ and $26\%$, respectively. Higher gains can be further obtained by increasing the storage capability at the network edge. 
\end{abstract}
%
%%%%%%%%%%%%%%%%%%%%%%%%%%%%%%%%%%%%%%%%%%%%%%%%%%%%%%%%%%%%%%%%%%%
%%%%%%%%%%%%%%%%%%%%%%%%%%%%%%%%%% INTRODUCTION
\section{Introduction}
\begin{figure*}[t]
	\centering
	\includegraphics[width=1.0\textwidth]{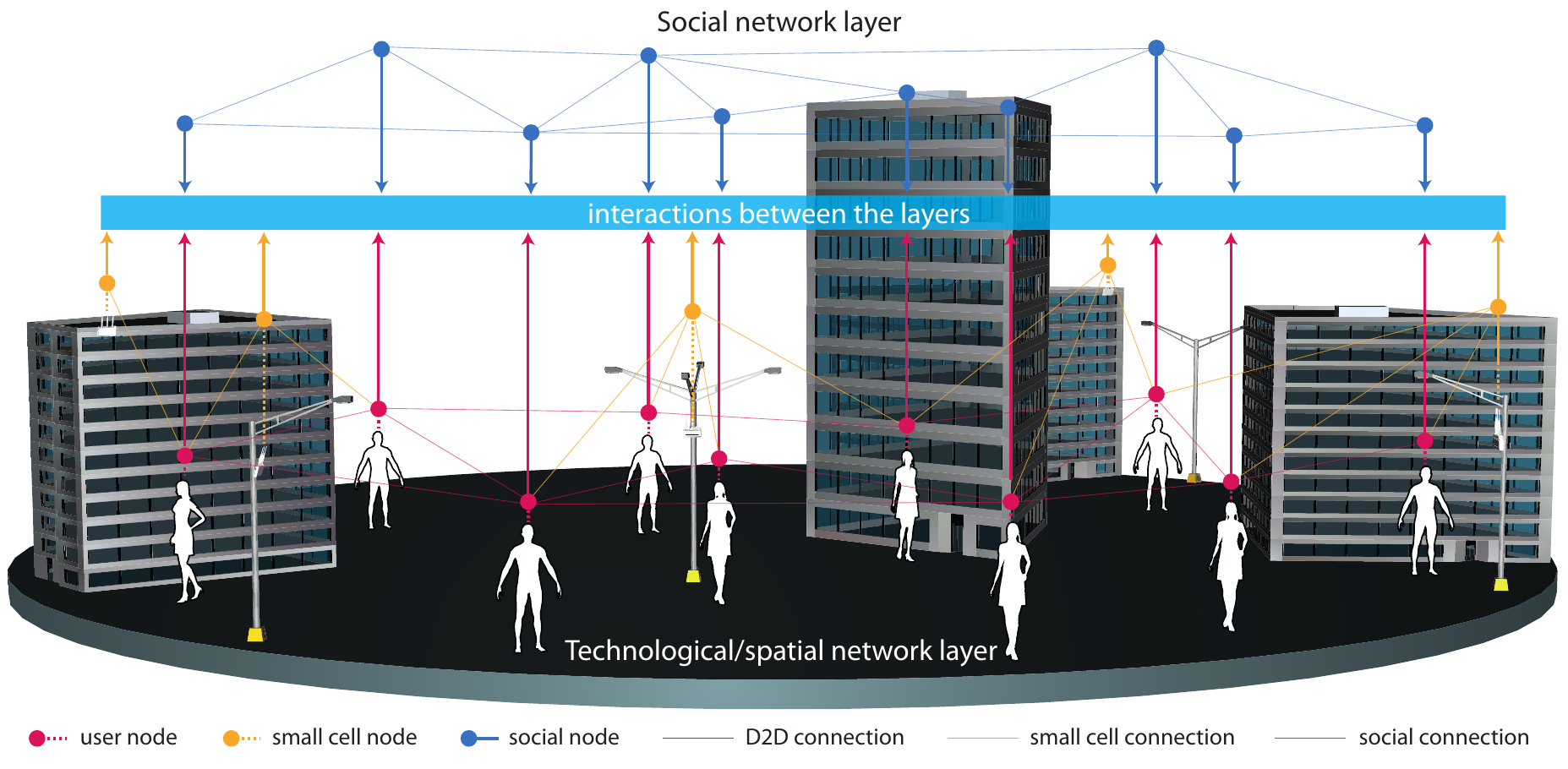}
	\caption{An illustration of an overlay of socially-interconnected and technological/spatial network.}
	\label{fig:scenario}
\end{figure*}
The recent proliferation of smartphones has substantially enriched the mobile user experience, leading to a vast array of new wireless services, including multimedia streaming, web-browsing applications and socially-interconnected networks. This phenomenon has been further fueled by mobile video streaming, which currently accounts for almost 50$\%$ of mobile data traffic, with a projection of 500-fold increase over the next 10 years \cite{cisco1}. At the same time, social networking is already the second largest traffic volume contributor with  a 15$\%$ average share \cite{Ericsson}. This new phenomenon has urged mobile operators to redesign their current networks and seek more advanced and sophisticated techniques to increase coverage, boost network capacity, and cost-effectively  bring contents closer to users.

A promising approach to meet these unprecedented traffic demands is via the deployment of \glspl{SCN} \cite{Jeff}. \glspl{SCN} represent a novel networking paradigm based on the idea of deploying short-range, low-power, and low-cost \glspl{SBS} underlaying the macrocellular network. To date, the vast majority of research works has been dealing with issues related to self-organization, \ac{ICIC}, traffic offloading, energy-efficiency, etc (see \cite{Jeff} and references therein). These studies were carried out under the existing \emph{reactive} networking paradigm, in which users' traffic requests and flows must be served urgently upon their arrival or dropped causing outages. Because of this, the existing small cell networking paradigm falls short of solving  peak traffic demands whose large-scale deployment hinges on expensive site acquisition, installation and backhaul costs. These shortcomings are set to become increasingly acute, due to the surging number of connected devices and the advent of ultra-dense networks, which will continue to strain current cellular network infrastructures. These key observations mandate a \emph{novel} networking paradigm which goes beyond current heterogeneous small cell deployments leveraging the latest developments in storage, context-awareness, and social networking \cite{intel}.

The proposed networking paradigm is proactive in essence and is rooted in the fact that  network nodes (i.e., base stations and handhelds/smartphones) exploit users' context information, anticipate users' demands and leverage their predictive abilities to achieve significant resource savings  to guarantee \ac{QoS} requirements and cost/energy expenditures \cite{ejder}. This paradigm goes beyond present cellular deployments, which have been designed assuming \emph{dumb} devices with very limited storage and processing power. Nevertheless, current smartphones have become very sophisticated devices with enhanced computing and storage capabilities. As a result, under the proactive networking paradigm, network nodes track, learn and build users' demand profiles to predict future requests, leveraging  devices' capabilities  and the vast amount of available data. Recently, predictive analytics and big data have received significant attention using  machine learning techniques to ingest and analyze mountains of infrastructure logs to produce predictive and actionable information for outage prediction and content recommendation \cite{mdc}. Endowed with these predictive capabilities, users are scheduled in a more efficient manner and resources are pre-allocated more intelligently, by proactively serving predictable peak-hour demands during off-peak times (for e.g., at night). By smartly exploiting the statistical traffic patterns and users' context information (i.e., file popularity distributions, location, velocity and mobility patterns), the proposed paradigm allows to better predict when users' contents are requested with the amount of resources needed, and at which network locations should contents be pre-cached.

Another topical trend is online \emph{social networks} (i.e., Facebook, Twitter, Digg) which have become instrumental in users' content distribution \cite{Ericsson}. As a matter of fact, users tend to value highly recommended contents by friends or people with similar interests and are also likely to recommend it. Thus, exploiting humans' interdependence through users' social relationships and ties, future networks can learn correlation patterns in networks of linked social and geographic data for a better prediction and inference of users' behavior. Fig. \ref{fig:scenario} shows an abstraction of the technological/spatial network layer overlaid with the social network layer. Since content dissemination of the nodes in social network layer is handled in real via the nodes in technological/spatial network layer, analyzing interactions between these two layers would yield further gains in future networks.
%
%%%%%%%%%%%%%%%%%%%%%%%%%%%%%%%%%% Prior Work
\subsection{Prior Work and Our Contribution}
The idea of femtocaching was proposed in which \glspl{SBS} have low-bandwidth (possibly wireless) backhaul links and high storage capabilities \cite{caire3}. The work in \cite{gamal} explored the notion of proactive resource allocation exploiting the predictability of user behavior for load balancing. Therein, using tools from large deviation theory, the scaling law of the outage probability is derived as a function of a prediction time window. Similarly, \cite{caire} studied the asymptotic scaling laws of caching in \ac{D2D} in which users collaborate by caching popular content and utilizing  \ac{D2D} communication. Nevertheless, while interesting, these works do not deal with the dynamics of proactive caching, overlooking aspects of context-awareness and social networks. These key aspects precisely constitute the prime motivations of this article, whose aim is to fill the void in the dynamics of proactive network caching.

The rest of this article is organized as follows. In Section \ref{sec:reactive2proactive}, a discussion of the limitations and challenges of current reactive \ac{SCN} deployments is discussed. In Sections \ref{sec:casestudy1}-\ref{sec:casestudy2}, the novel proactive caching paradigm and its key ingredients are described. In addition,  two  case studies are presented to show the effectiveness of proactive caching. Finally, Section \ref{sec:conclusions} draws conclusions and future work.
%
%%%%%%%%%%%%%%%%%%%%%%%%%%%%%%%%%%%%%%%%%%%%%%%%%%%%%%%%%%%%%%%%%%%
%%%%%%%%%%%%%%%%%%%%%%%%%%%%%%%%%% FROM REACTIVE TO PROACTIVE NETWORKS
\section{From Reactive to Proactive Networks}
\label{sec:reactive2proactive}
The overarching goal of this article is to explore the foundations of small-cell enabled predictive/proactive \glspl{RAN}, and make a major leap forward on this novel networking paradigm. Cellular networks, increasingly, the most essential aspect of our telecommunication infrastructure, are in a period of unprecedented change, and hence incremental changes to current state-of-the-art for designing and optimizing such (reactive) networks are  becoming obsolete. The proposed framework rests on the notion that  network nodes anticipate users' demands and utilize their predictive abilities to reduce the traffic peak-to-average ratio, yielding significant  network resource savings. The  proactive approach leverages the existing heterogeneous cellular network and involves the design of predictive radio resource management techniques to maximize the efficiency of future $5$G  networks.
%%%%%%%%%%%%%%%%%%%%%%%%%%%%%%%%%% Leveraging Proactivity
\subsection{Leveraging Proactivity}
The predictive framework rests on the notion that  information demand patterns of mobile users are, to a certain extent, predictable. Such predictability can be exploited to minimize the peak load of cellular networks, by proactively pre-caching desired information to selected users before they actually request it. Leveraging the powerful processing capabilities and large memory storage of smart-phones enables network operators to proactively serve predictable peak-hour requests during off-peak times. That is, when the proactive network serves users' requests before their deadlines, the corresponding data is stored in the user device and, when the request is actually initiated, the information is pulled out directly from the cached memory instead of accessing the wireless network. For this purpose, novel machine learning techniques should be developed to find optimal tradeoffs between predictions that result in content being retrieved that  users ultimately never request and requests not anticipated in a timely manner. Clearly, analyzing user's traffic and caching content locally at the \ac{SBS} and user terminal can significantly reduce the backhaul traffic, notably when networks are inundated with similar requests for content. Hence, the objective is to predict, anticipate, and infer on future events in an intelligent manner, which is a complex problem exacerbated by the \emph{big data} paradigm induced by the large and sparse information/data \cite{bigdata}. Indeed, data sparsity is a key challenge since it may not be always possible to collect enough data from a single user   to predict her/his patterns precisely enough. To overcome this challenge, other users' data as well as their social relationships can be leveraged to build reliable statistical models. Of paramount importance is over a time window which contents should \glspl{SBS} pre-allocate? When (at which time slot should it be pre-scheduled)? To which strategic/influential users? And in which location in the network?.
%%%%%%%%%%%%%%%%%%%%%%%%%%%%%%%%%% Leveraging Social Networks
\subsection{Leveraging Social Networks}
Yet, another untapped paradigm of beyond $4$G networks to provide unlimited access to information for anyone and anything, is undoubtedly social networks. Indeed, social networks are redefining the way data is accessed throughout the network, exploiting  social relationships and ties among users, to better optimize network resources. Harnessing how users encounter each other within their social communities, local \ac{D2D} communication is key in pre-allocating strategic contents in the caches of important/influential users.

Driven by the fact that the volume of mobile data will be 1000X higher than today, and between 10 to 100X more connected devices by $2020$, future networks will need to manage a massive amount of connected devices \cite{cisco1}. In fact, already today, the vast majority of data traffic is carried out by social networks, which have played a crucial role in information propagation over the Internet, and will continue to shape up the way information is accessed. The social characteristics such as the external influence from media and friends, users' relationships and ties can help better plan future networks. In particular, by exploiting the correlation between users' data, their social interests and their common interests,  the accuracy of predicting future events (i.e., users' geographic positions, next visited cells, requested files) can be dramatically improved. For instance, \emph{geotagging} data in social networking applications can help operators track where people generate mobile data traffic to optimally deploy small cells. A by-product of this is helping operators in other aspects of network design such as: small cell handover, multi-tier interference management (since we know to which cell the user will connect next), power management and greener networks by serving users only when close to the small cell.

In the next section, we show the benefits and prospects of proactive networking via two  case studies, leveraging \ac{SCN} deployments and notions of machine learning and social networks.
%
%%%%%%%%%%%%%%%%%%%%%%%%%%%%%%%%%%%%%%%%%%%%%%%%%%%%%%%%%%%%%%%%%%%
%%%%%%%%%%%%%%%%%%%%%%%%%%%%%%%%%% CASE STUDY I
\section{Case Study I: Proactive Small Cell Networks}
\label{sec:casestudy1}
In this section, we investigate the problem of backhaul offloading in \glspl{SCN}, in which proactive caching plays a crucial role. Indeed, backhauling is of utmost importance before a roll-out of \glspl{SCN}. In the considered network model, \glspl{SBS} are deployed with high capacity storage units but have limited capacity backhaul links. We build on \cite{ejder}, in which a proactive caching procedure is proposed to store files based on their highest popularity, until the storage capacity is achieved. Therein,  \glspl{SBS}    have perfect information of the popularity matrix $\bold{P}_{N \times F}$ where each row represents users and columns file preferences/ratings. Nevertheless, in practice, the popularity matrix is  large, sparse and partially unknown. Therefore,  inspired from the \emph{Netflix paradigm} and using tools from supervised machine learning and specifically \ac{CF}, we propose a distributed proactive caching procedure that exploits  users-files correlations to infer on the probability that the $u$-th user requests the $i$-th file. 

The proposed caching procedure is composed of a training and placement part. In the training part, the goal is to estimate the popularity matrix ${\bf P}$ (namely ${\hat{\bf P}}_{N \times F}$), where every \ac{SBS} builds a model based on the already available information regarding users' preferences/ratings\footnote{Depending on the operator's choice and load conditions of the \glspl{SBS}, the training part can be done in a central unit instead of \glspl{SBS}.}. This is done by solving the following least square minimization problem:
\begin{equation}
\min_{\{b_u,b_i\}} \sum_{u,i} \Big(r_{ui}-\hat{r}_{ui}\Big)^2+\lambda \Big(\sum_ {u}b_u^2+\sum_ {i}b_i^2\Big)
\end{equation}
where the sum is  over the $(u, i)$ user/file pairs in the training set where user $u$ actually rated file $i$ (i.e., $r_{ui}$), and the minimization is over  the $N + F$ parameters, where $N$ is the number of users and $F$ the number of files in the training set. In addition, $\hat{r}_{ui}=\bar{r}+b_u+b_i$ is the baseline predictor in which  $b_ i$ models the quality of each file $i$ relative to the average $\bar{r}$, and $b_u$ models the quality of each user $u$ relative to $\bar{r}$. Finally, the weight $\lambda$ is chosen to balance between regularization and fitting training data. In the experimental setup, the regularized \ac{SVD}  was used for its numerical accuracy (see \cite{CF} for other \ac{CF} methods and their comparison). Regularized \ac{SVD} based \ac{CF} constructs $\bold{\hat{P}}$, as  the low rank version of $\bold{P}$. Since the training set is sparse, the decomposition is done via gradient descent by exploiting the least-squares property of \ac{SVD}. After obtaining the estimated file popularity matrix $\bold{\hat{P}}$, the proactive caching decision can be made in the placement phase by storing the most popular files greedily (as in \cite{ejder}) until no storage space remains.
%
%%%%%%%%%%%%%%%%%%%%%%%%%%%%%%%%%% Case Study 1: Numerical results and discussion
\subsection{Numerical results and discussion}
The experimental setup for the proactive caching procedure includes $M$ SBSs and $N$ users. The sum capacity of the wireless links between the \glspl{SBS} and users is $C_w$. For simplification, these link/storage capacities are assumed to be equal. File requests of users are drawn from a  library of size $F$, where each file $f_i$ has length $L$ and bitrate requirement $B$. A user's request is said to be \emph{satisfied} if the   delivery duration is below a certain threshold, which is a function of the bitrate of the requested file. The \emph{backhaul load} is defined as the amount of bandwidth consumed by the backhaul links over the wireless bandwidth. The list of parameters is given in Table \ref{tab:setup_params1}. In the simulations, we consider two regimes of interest: (i) low load and (ii) high load.

For a given number of requests $R$ and time duration $T$, the arrival times of requesting users are drawn uniformly at random, and the files' samples are obtained from the $\textrm{ZipF}(\alpha)$ distribution\footnote{Evidence of such a distribution is observed in many real-world phenomena including distributions of files in the web proxies \cite{zipf}. Briefly, $\alpha$ is the exponent characterizing the ZipF distribution in which $\alpha \rightarrow \infty$ implies a steeper distribution whereas $\alpha \rightarrow 0$ makes the distribution more uniform.}. At time instant $t=0$, the perfect popularity matrix is constructed out of which 20$\%$ of the elements are removed uniformly at random and the remaining matrix is used for training. The  removed entries are predicted using the Regularized SVD \cite{CF} and the estimated matrix is then used in the proactive caching procedure by storing these popular files under  storage constraints. The precaching decision is carried out by each \ac{SBS} until all requests are served. For comparison purposes and to mimic the reactive scenario, random caching is used as a baseline.

For the performance curves, three different parameters of interest are considered: (i) number of requests $R$, (ii) cache size $S$, and (iii) ZipF distribution parameter $\alpha$. To show the percentages of differences between the proactive and  reactive approaches, the  number of requests are normalized by ${R^{\star}}$, cache size by $L \times F$, and $\alpha$ by $2$. These normalized parameters are denoted by ${\widehat R}$, ${\widehat S}$ and ${\widehat \alpha}$ respectively. The performance of the number of satisfied requests and backhaul loads are shown in Fig. \ref{fig:case1}. Each figure represents the variation of one parameter while the rest is fixed for different regimes.
%
%%%%%%%%%%%%%%%%%%%%%%%%%%%%%%%%%% Table 1 - Fixed/scaled parameters for case study 1
\begin{table}[ht]
\centering
\begin{tabular}{|c|l|c|}
\hline
\textbf{Parameter}			  &  \textbf{Description} 	& 	\textbf{Value}\\
\hline
$T$    		& Time slots							& 	$ 1024$ seconds \\
\hline
$M$   		& Number of small cells 				& 	$4$ \\
\hline
$N$ 			& Number of user terminals 			& 	$32$ 	\\
\hline
$F$			& Number of files 					&  $128$	 \\
\hline
$L$  		& Length of each file 				& 	$1$	Mbit \\
\hline
$B$ 			& Bitrate of each file				&	$1$ Mbit/s	\\
\hline
$C_b$ 		& Total backhaul link capacity		&	$2$	Mbit/s	\\
\hline
$C_w$ 		& Total wireless link capacity 		&	$64$	 Mbit/s\\
\hline
\hline
$R^{\star}$	& 	Maximum number of requests 		&	$2048$ 	\\
\hline
\hline
$R$ 			&	Number of requests 				&		$0 \sim 2048$\\
\hline
$S$    		&  Cache size							& 	$0 \sim 128$ Mbit\\
\hline
$\alpha$   	& 	ZipF parameter	 					& $0 \sim 2$\\
\hline
\end{tabular}
\vspace{3mm}
\caption{List of parameters for case study I.}
\label{tab:setup_params1}
\end{table}
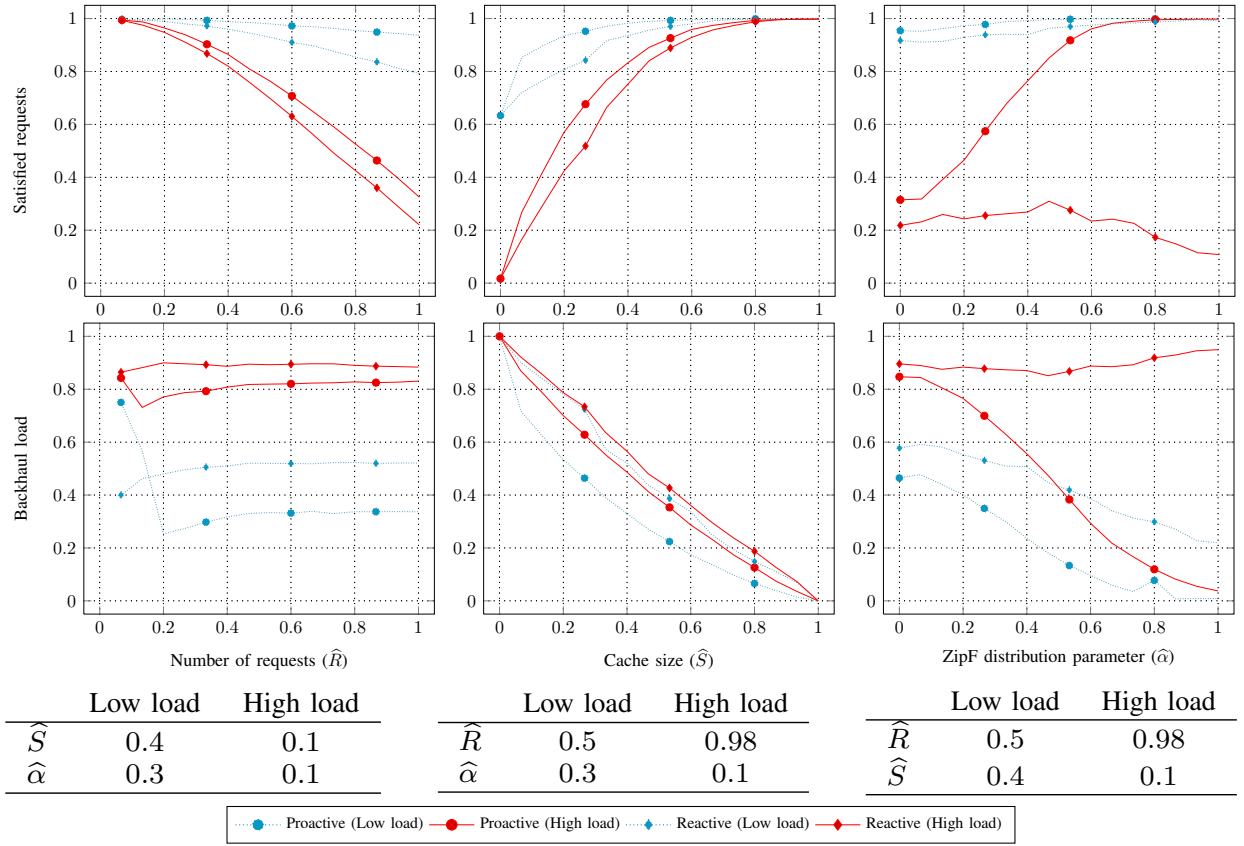
\begin{figure*}[!ht]
\centering
%%%%%%%%%%%%%%%%%%%%%%%%%%%%%%%%%%%% Plots Case 1: Satisfied requests
\begin{tikzpicture}[scale=0.68, baseline]
	\begin{axis}[
		grid = major,
		cycle list name=thechairLivingOnTheEdge,
		mark repeat={4},
		ymin=-0.05,ymax=1.05, ytick={0, 0.2, 0.4, 0.6, 0.8, 1},
		xmin=-0.05,xmax=1.05, xtick={0, 0.2, 0.4, 0.6, 0.8, 1},
	  legend columns=4,
		legend entries={Proactive (Low load), Proactive (High load), Reactive (Low load), Reactive (High load)},
		legend cell align=left,
	  legend style ={font=\tiny},
		legend to name=namedmethods,
		ylabel =Satisfied requests]
		
	\addplot plot coordinates {
		(0.066667,	0.995)
		(0.13333,	0.99341)
		(0.2,	0.99905)
		(0.26667,	0.99669)
		(0.33333,	0.99302)
		(0.4,	0.98839)
		(0.46667,	0.98497)
		(0.53333,	0.97969)
		(0.6,	0.97236)
		(0.66667,	0.96883)
		(0.73333,	0.96262)
		(0.8,	0.95479)
		(0.86667,	0.9489)
		(0.93333,	0.94364)
		(1,	0.93661)
	}; 	

	\addplot plot coordinates {
		(0.066667,	0.99436)
		(0.13333,	0.98695)
		(0.2,	0.96446)
		(0.26667,	0.93807)
		(0.33333,	0.90279)
		(0.4,	0.86425)
		(0.46667,	0.80953)
		(0.53333,	0.76215)
		(0.6,	0.70738)
		(0.66667,	0.648)
		(0.73333,	0.58876)
		(0.8,	0.52506)
		(0.86667,	0.46347)
		(0.93333,	0.39607)
		(1,	0.32488)
	}; 	

	\addplot plot coordinates {
		(0.066667,	0.99779)
		(0.13333,	0.99498)
		(0.2,	0.98926)
		(0.26667,	0.9795)
		(0.33333,	0.97207)
		(0.4,	0.95973)
		(0.46667,	0.94548)
		(0.53333,	0.92897)
		(0.6,	0.91008)
		(0.66667,	0.89735)
		(0.73333,	0.87526)
		(0.8,	0.85463)
		(0.86667,	0.83582)
		(0.93333,	0.81265)
		(1,	0.79273)
	}; 	

	\addplot plot coordinates {
		(0.066667,	0.99307)
		(0.13333,	0.9747)
		(0.2,	0.94778)
		(0.26667,	0.9097)
		(0.33333,	0.86761)
		(0.4,	0.82013)
		(0.46667,	0.75907)
		(0.53333,	0.69629)
		(0.6,	0.63029)
		(0.66667,	0.56122)
		(0.73333,	0.48944)
		(0.8,	0.425)
		(0.86667,	0.36001)
		(0.93333,	0.28916)
		(1,	0.21973)
	}; 	

	\end{axis}
\end{tikzpicture}
%%%%%%%%%%%%%%%%%%
\begin{tikzpicture}[scale=0.68, baseline]
	\begin{axis}[
		grid = major,
		cycle list name=thechairLivingOnTheEdge,
		ymin=-0.05,ymax=1.05, ytick={0, 0.2, 0.4, 0.6, 0.8, 1},
		xmin=-0.05,xmax=1.05, xtick={0, 0.2, 0.4, 0.6, 0.8, 1},
		mark repeat={4}]
		
	\addplot plot coordinates {
		(0,	0.63333)
		(0.066667,	0.85313)
		(0.13333,	0.8976)
		(0.2,	0.93511)
		(0.26667,	0.95189)
		(0.33333,	0.97119)
		(0.4,	0.98241)
		(0.46667,	0.98962)
		(0.53333,	0.99309)
		(0.6,	0.99591)
		(0.66667,	0.99836)
		(0.73333,	0.99911)
		(0.8,	0.99917)
		(0.86667,	0.99946)
		(0.93333,	0.99946)
		(1,	0.99946)
	}; 	

	\addplot plot coordinates {
		(0,	0.016821)
		(0.066667,	0.26746)
		(0.13333,	0.42249)
		(0.2,	0.57112)
		(0.26667,	0.6763)
		(0.33333,	0.76794)
		(0.4,	0.83307)
		(0.46667,	0.89136)
		(0.53333,	0.92602)
		(0.6,	0.95889)
		(0.66667,	0.97429)
		(0.73333,	0.98495)
		(0.8,	0.99306)
		(0.86667,	0.99574)
		(0.93333,	0.99738)
		(1,	0.99792)
	}; 	

	\addplot plot coordinates {
		(0,	0.63333)
		(0.066667,	0.72078)
		(0.13333,	0.76615)
		(0.2,	0.80756)
		(0.26667,	0.84215)
		(0.33333,	0.91706)
		(0.4,	0.93564)
		(0.46667,	0.95618)
		(0.53333,	0.97033)
		(0.6,	0.97968)
		(0.66667,	0.99131)
		(0.73333,	0.99669)
		(0.8,	0.9981)
		(0.86667,	0.99873)
		(0.93333,	0.9994)
		(1,	0.99946)
	}; 	

	\addplot plot coordinates {
		(0,	0.016821)
		(0.066667,	0.16533)
		(0.13333,	0.29582)
		(0.2,	0.42469)
		(0.26667,	0.51813)
		(0.33333,	0.6647)
		(0.4,	0.75219)
		(0.46667,	0.84089)
		(0.53333,	0.88881)
		(0.6,	0.92889)
		(0.66667,	0.95727)
		(0.73333,	0.97409)
		(0.8,	0.98774)
		(0.86667,	0.99544)
		(0.93333,	0.99739)
		(1,	0.99792)
	}; 	

	\end{axis}
\end{tikzpicture}
%%%%%%%%%%%%%%%%%%
\begin{tikzpicture}[scale=0.68, baseline]
	\begin{axis}[
		grid = major,
		cycle list name=thechairLivingOnTheEdge,
		ymin=-0.05,ymax=1.05, ytick={0, 0.2, 0.4, 0.6, 0.8, 1},
		xmin=-0.05,xmax=1.05, xtick={0, 0.2, 0.4, 0.6, 0.8, 1},
		mark repeat={4}]
		
	\addplot plot coordinates {
		(0,	0.95417)
		(0.066667,	0.952)
		(0.13333,	0.96219)
		(0.2,	0.97292)
		(0.26667,	0.97766)
		(0.33333,	0.98889)
		(0.4,	0.99217)
		(0.46667,	0.99677)
		(0.53333,	0.99743)
		(0.6,	0.99885)
		(0.66667,	0.99943)
		(0.73333,	0.99946)
		(0.8,	0.99934)
		(0.86667,	0.99946)
		(0.93333,	0.99946)
		(1,	0.99946)
	}; 	

	\addplot plot coordinates {
		(0,	0.31472)
		(0.066667,	0.31788)
		(0.13333,	0.39238)
		(0.2,	0.46472)
		(0.26667,	0.5742)
		(0.33333,	0.67842)
		(0.4,	0.76458)
		(0.46667,	0.85102)
		(0.53333,	0.91777)
		(0.6,	0.96116)
		(0.66667,	0.98223)
		(0.73333,	0.99012)
		(0.8,	0.9956)
		(0.86667,	0.99675)
		(0.93333,	0.99757)
		(1,	0.99792)
	}; 	

	\addplot plot coordinates {
		(0,	0.9171)
		(0.066667,	0.91095)
		(0.13333,	0.91388)
		(0.2,	0.92874)
		(0.26667,	0.93831)
		(0.33333,	0.94082)
		(0.4,	0.93941)
		(0.46667,	0.96325)
		(0.53333,	0.96983)
		(0.6,	0.9754)
		(0.66667,	0.98008)
		(0.73333,	0.98382)
		(0.8,	0.98696)
		(0.86667,	0.99087)
		(0.93333,	0.99538)
		(1,	0.99169)
	}; 	

	\addplot plot coordinates {
		(0,	0.21808)
		(0.066667,	0.23178)
		(0.13333,	0.25951)
		(0.2,	0.2427)
		(0.26667,	0.25538)
		(0.33333,	0.26205)
		(0.4,	0.26877)
		(0.46667,	0.3093)
		(0.53333,	0.27571)
		(0.6,	0.23449)
		(0.66667,	0.24185)
		(0.73333,	0.22582)
		(0.8,	0.17312)
		(0.86667,	0.14743)
		(0.93333,	0.11469)
		(1,	0.10799)
	}; 	

	\end{axis}
\end{tikzpicture}
%%%%%%%%%%%%%%%%%%%%%%%%%%%%%%%%%%%% Plots Case 1: Backhaul load
\begin{tikzpicture}[scale=0.68, baseline]
	\begin{axis}[
		grid = major,
		cycle list name=thechairLivingOnTheEdge,
		mark repeat={4},
		ymin=-0.05,ymax=1.05, ytick={0, 0.2, 0.4, 0.6, 0.8, 1},
		xmin=-0.05,xmax=1.05, xtick={0, 0.2, 0.4, 0.6, 0.8, 1},
		xlabel=Number of requests ($\widehat{R}$),
		ylabel=Backhaul load]
		
	\addplot plot coordinates {
		(0.066667,	0.75)
		(0.13333,	0.56645)
		(0.2,	0.25298)
		(0.26667,	0.2737)
		(0.33333,	0.29757)
		(0.4,	0.31742)
		(0.46667,	0.33027)
		(0.53333,	0.33362)
		(0.6,	0.33147)
		(0.66667,	0.33858)
		(0.73333,	0.32986)
		(0.8,	0.33695)
		(0.86667,	0.33695)
		(0.93333,	0.33742)
		(1,	0.33683)
	};	

	\addplot plot coordinates {
		(0.066667,	0.84286)
		(0.13333,	0.73118)
		(0.2,	0.77088)
		(0.26667,	0.78712)
		(0.33333,	0.79256)
		(0.4,	0.80788)
		(0.46667,	0.818)
		(0.53333,	0.81932)
		(0.6,	0.82035)
		(0.66667,	0.82319)
		(0.73333,	0.82433)
		(0.8,	0.82767)
		(0.86667,	0.82499)
		(0.93333,	0.82669)
		(1,	0.83063)
	};	

	\addplot plot coordinates {
		(0.066667,	0.4)
		(0.13333,	0.46237)
		(0.2,	0.47971)
		(0.26667,	0.49553)
		(0.33333,	0.50501)
		(0.4,	0.50955)
		(0.46667,	0.52045)
		(0.53333,	0.51968)
		(0.6,	0.51908)
		(0.66667,	0.51825)
		(0.73333,	0.52245)
		(0.8,	0.52296)
		(0.86667,	0.51954)
		(0.93333,	0.52096)
		(1,	0.52099)
	};	

	\addplot plot coordinates {
		(0.066667,	0.86429)
		(0.13333,	0.88172)
		(0.2,	0.89976)
		(0.26667,	0.89624)
		(0.33333,	0.8927)
		(0.4,	0.88663)
		(0.46667,	0.89468)
		(0.53333,	0.89267)
		(0.6,	0.89428)
		(0.66667,	0.89621)
		(0.73333,	0.8959)
		(0.8,	0.89028)
		(0.86667,	0.88718)
		(0.93333,	0.88548)
		(1,	0.88359)
	};	

	\end{axis}
\end{tikzpicture}
%%%%%%%%%%%%%%%%%%
\begin{tikzpicture}[scale=0.68, baseline]
	\begin{axis}[
		grid = major,
		cycle list name=thechairLivingOnTheEdge,
		ymin=-0.05,ymax=1.05, ytick={0, 0.2, 0.4, 0.6, 0.8, 1},
		xmin=-0.05,xmax=1.05, xtick={0, 0.2, 0.4, 0.6, 0.8, 1},
		mark repeat={4},
		xlabel=Cache size ($\widehat{S}$)]
		
	\addplot plot coordinates {
		(0,	1)
		(0.066667,	0.71851)
		(0.13333,	0.62405)
		(0.2,	0.5334)
		(0.26667,	0.46374)
		(0.33333,	0.38931)
		(0.4,	0.33015)
		(0.46667,	0.27099)
		(0.53333,	0.22399)
		(0.6,	0.17465)
		(0.66667,	0.13645)
		(0.73333,	0.098282)
		(0.8,	0.06584)
		(0.86667,	0.039132)
		(0.93333,	0.014313)
		(1,	0)
	};	

	\addplot plot coordinates {
		(0,	1)
		(0.066667,	0.86904)
		(0.13333,	0.78773)
		(0.2,	0.7001)
		(0.26667,	0.62804)
		(0.33333,	0.55161)
		(0.4,	0.48637)
		(0.46667,	0.41285)
		(0.53333,	0.35346)
		(0.6,	0.28676)
		(0.66667,	0.23272)
		(0.73333,	0.17462)
		(0.8,	0.1261)
		(0.86667,	0.074976)
		(0.93333,	0.03554)
		(1,	0)
	};	

	\addplot plot coordinates {
		(0,	1)
		(0.066667,	0.90076)
		(0.13333,	0.83969)
		(0.2,	0.78149)
		(0.26667,	0.72424)
		(0.33333,	0.57156)
		(0.4,	0.52004)
		(0.46667,	0.43798)
		(0.53333,	0.38645)
		(0.6,	0.33683)
		(0.66667,	0.24905)
		(0.73333,	0.19466)
		(0.8,	0.14885)
		(0.86667,	0.11069)
		(0.93333,	0.068702)
		(1,	0)
	};	

	\addplot plot coordinates {
		(0,	1)
		(0.066667,	0.92162)
		(0.13333,	0.85686)
		(0.2,	0.78724)
		(0.26667,	0.73369)
		(0.33333,	0.63535)
		(0.4,	0.56524)
		(0.46667,	0.48004)
		(0.53333,	0.42697)
		(0.6,	0.36076)
		(0.66667,	0.29601)
		(0.73333,	0.23856)
		(0.8,	0.18744)
		(0.86667,	0.12707)
		(0.93333,	0.072541)
		(1,	0)
	};	

	\end{axis}
\end{tikzpicture}
%%%%%%%%%%%%%%%%%%
\begin{tikzpicture}[scale=0.68, baseline]
	\begin{axis}[
		grid = major,
		cycle list name=thechairLivingOnTheEdge,
		ymin=-0.05,ymax=1.05, ytick={0, 0.2, 0.4, 0.6, 0.8, 1},
		xmin=-0.05,xmax=1.05, xtick={0, 0.2, 0.4, 0.6, 0.8, 1},		
		mark repeat={4},
		xlabel=ZipF distribution parameter ($\widehat{\alpha}$)]
		
	\addplot plot coordinates {
		(0,	0.46469)
		(0.066667,	0.4771)
		(0.13333,	0.43989)
		(0.2,	0.40076)
		(0.26667,	0.34934)
		(0.33333,	0.29975)
		(0.4,	0.23569)
		(0.46667,	0.1813)
		(0.53333,	0.13358)
		(0.6,	0.096374)
		(0.66667,	0.05916)
		(0.73333,	0.035305)
		(0.8,	0.07729)
		(0.86667,	0.0076336)
		(0.93333,	0.0085878)
		(1,	0.0085878)
	};	

	\addplot plot coordinates {
		(0,	0.84713)
		(0.066667,	0.84421)
		(0.13333,	0.80526)
		(0.2,	0.76534)
		(0.26667,	0.69961)
		(0.33333,	0.62999)
		(0.4,	0.55599)
		(0.46667,	0.4742)
		(0.53333,	0.38315)
		(0.6,	0.29309)
		(0.66667,	0.21811)
		(0.73333,	0.16699)
		(0.8,	0.11928)
		(0.86667,	0.081792)
		(0.93333,	0.055015)
		(1,	0.037488)
	};	

	\addplot plot coordinates {
		(0,	0.57729)
		(0.066667,	0.5916)
		(0.13333,	0.58111)
		(0.2,	0.55248)
		(0.26667,	0.53053)
		(0.33333,	0.50954)
		(0.4,	0.50763)
		(0.46667,	0.44656)
		(0.53333,	0.41889)
		(0.6,	0.3874)
		(0.66667,	0.34065)
		(0.73333,	0.31298)
		(0.8,	0.29866)
		(0.86667,	0.27099)
		(0.93333,	0.2271)
		(1,	0.21947)
	};	

	\addplot plot coordinates {
		(0,	0.89581)
		(0.066667,	0.88948)
		(0.13333,	0.87537)
		(0.2,	0.88364)
		(0.26667,	0.8778)
		(0.33333,	0.87342)
		(0.4,	0.8705)
		(0.46667,	0.85102)
		(0.53333,	0.86758)
		(0.6,	0.88754)
		(0.66667,	0.8851)
		(0.73333,	0.89241)
		(0.8,	0.9187)
		(0.86667,	0.92941)
		(0.93333,	0.94547)
		(1,	0.94937)
	};	

	\end{axis}
\end{tikzpicture}
%%%%%%%%%%%%%%%%%%%%%%%%%%%%%%%%%%%% Parameters
\begin{tikzpicture}[scale=0.68, baseline]
\node (table) {
\scriptsize
\resizebox{0.29\linewidth}{!}{
\begin{tabular}{l*{3}{c}}
& Low load & High load \\
\hline
${\widehat S}$ 			& $0.4$ & $0.1$ \\
${\widehat \alpha}$ & $0.3$ & $0.1$ \\
\hline
\end{tabular}
}
};
\end{tikzpicture}
%%%%%%%%%%%%%%%%%%
\begin{tikzpicture}[scale=0.68, baseline]
\node (table) {
\resizebox{0.29\linewidth}{!}{
\scriptsize
\begin{tabular}{l*{3}{c}}
& Low load & High load \\
\hline
${\widehat R}$ 			& $0.5$ & $0.98 $ \\
${\widehat \alpha}$ 		& $0.3$ & $0.1$ \\
\hline
\end{tabular}
}
};
\end{tikzpicture}
%%%%%%%%%%%%%%%%%%
\begin{tikzpicture}[scale=0.68, baseline]
\node (table) {
\resizebox{0.29\linewidth}{!}{
\scriptsize
\begin{tabular}{l*{3}{c}}
& Low load &  High load \\
\hline
${\widehat R}$ 			& $0.5$ & $0.98 $ \\
${\widehat S}$ 			& $0.4$ & $0.1$ \\
\hline
\end{tabular}
}
};
\end{tikzpicture}
\\
\ref{namedmethods}
\caption{Proactive Small Cell Networks: Evolutions of satisfied requests and backhaul load with respect to number of requests, cache size and ZipF parameter.}
\label{fig:case1}
\end{figure*} 
\subsubsection{Impact of number of requests}
As the number of users' requests  increases, the amount of satisfied requests starts decreasing due to the limited resource constraints. However, the proactive caching approach outperforms the reactive one in terms of satisfied requests. On the other hand, for very small users' requests, the reactive approach  generates  less load on the backhaul. This situation is due to the \emph{cold start} phenomena in which  \ac{CF} cannot draw any inference due to non-sufficient amount of information about the popularity matrix. Hence, caching randomly from a fixed library may relatively perform better under very low loads. However, as users’ requests increase the proactive approach tends to decrease the backhaul load  outperforming the reactive approach. The gains become constant after a certain point.
\subsubsection{Impact of cache size}
As ${\widehat S}$ increases, the number of satisfactions approaches $1$ and the backhaul load  becomes $0$. This reflects the unrealistic case where all requested files can be cached. Assuming this is not the case in reality and checking for intermediate values of cache sizes, it can be seen that  proactive caching outperforms the reactive case.
\subsubsection{Impact of popularity distribution}
As some files  become more popular than others (${\widehat \alpha}$ increases), the gain between  proactive and reactive caching  is higher in all load regimes. In addition, the gains further increase with higher incoming loads both in terms of satisfied requests and backhaul load.
%
%%%%%%%%%%%%%%%%%%%%%%%%%%%%%%%%%%%%%%%%%%%%%%%%%%%%%%%%%%%%%%%%%%%
%%%%%%%%%%%%%%%%%%%%%%%%%%%%%%%%%% CASE STUDY II
\section{Case Study II: Social Networks Aware Caching via D2D}
\label{sec:casestudy2}
In this section, we show the effectiveness of proactive caching leveraging social networks and \ac{D2D} communications. Specifically, we consider a network deployment where users seek certain files from a given library of $F$ files. Each user can store files on its device subject to its storage capacity. As shown in Fig. \ref{fig:scenario}, the considered network can be viewed as an overlay of both social and small cell network.

By exploiting the interplay between social and technological networking, each \ac{SBS} tracks and learns the set of \emph{influential} users using the social graph, and determines the influence probabilities based on past action history of users' encounters and file requests. Notably, when a given user requests a particular file, the \ac{SBS} determines whether one of the influential users has the requested file. If so, it directs the influential user to communicate the file to the requesting user via \ac{D2D}. Otherwise, if the file is not cached by the influential user, the \ac{SBS} transmits the file directly to the requesting user from the core network.

In order to determine the set of influential users, we exploit the social relationships and ties among users using the notion of \emph{centrality} metric \cite{NetworksNewman}. The centrality metric measures the social influence of a node based on how well it connects the network, whereby a node with higher centrality is more important (i.e., influential) to its social community. Typically, four centrality metrics can be used: (1) \emph{degree centrality}, to represent the number of ties a node has with other nodes; (2) \emph{closeness centrality}, to represent the distance between a node and other nearby nodes. Besides, the closeness metric is key for capturing the most influential users; (3) \emph{betweenness centrality}, which represents the extent to which a node lies on the shortest paths linking to other nodes; (4) \emph{eigenvector centrality}, estimates influence of nodes in the network by using the eigenvector corresponding to the largest eigenvalue of the  adjacency matrix of the network. In this paper, the eigenvector centrality is used for detecting the set of influential users.
%
%%%%%%%%%%%%%%%%%%%%%%%%%%%%%%%%%% Social Community Formation
\subsection{Social Community Formation}
Let $G=(V,E)$ denote the corresponding social graph composed of $N$ nodes which can be completely described by the adjacency (or connectivity) matrix $\bold{A}_{N \times N}$ with entry $a_{ij}$, $i,j=1,...,N$ equals $1$ if link (or edge) $l_{ ij}$ exists,
or $0$ otherwise. Using one of the above-mentioned metrics (i.e., centrality, closeness, and betweenness) allows us to describe the communication probability between two users, which can also be seen as the weight of the link between user $i$ and user $j$. Subsequently, knowing $\bold{A}$, each \ac{SBS} identifies the set of influential users which will be instrumental in proactively caching strategic contents\footnote{In practice, the computation and storage of $\bold{A}$ can be done in a central unit, in SBSs or in users terminals. Such a choice depends on the technical feasibility of detection and privacy concerns.}. Suppose that the eigenvalues of $\bold{A}$  are $\lambda_1, ..., \lambda_N$ in decreasing order and the corresponding eigenvectors are $\bold{v}_1, ... \bold{v}_N$. Then, eigenvector-centrality is basically the eigenvector $\bold{v}_1$ which corresponds to the largest eigenvalue that is $\lambda_1$. Thus, after obtaining the $K$-most influential users from $\bold{v}_1$, a clustering method can be applied for community formation.
%
%%%%%%%%%%%%%%%%%%%%%%%%%%%%%%%%%% Social-Aware Caching via D2D
\subsection{Social-Aware Caching via D2D}
After knowing the influential users and their communities, the next step is to determine the content dissemination process inside each community. For this purpose, we model the content dissemination as a \ac{CRP}, which is also known as a stochastic Dirichlet process. The prime motivation of using this process is to model the user-file partition procedure which essentially constitutes a prior information of how users match to files. Before going into details, we first define the number of users as $N$ and the total number of  contents by $F$. Given the large volume of contents available, we assume that $F = F_0+F_h$, in which $F_h$ represents the set of contents with viewing histories and $F_0$ is the set of contents without history. After the social communities have been formed, users seek their respective contents leveraging their social relationships and ties. We suppose that each user is interested in only one\footnote{The extension to the case of an arbitrary number of contents can be accommodated.} type of available contents $F$. Let $\pi_f$ denote the probability that content/file $f$ is selected by a given user, which we assume to follow a Beta distribution (i.e., prior) \cite{crp}. Thus, the selection result of user $n$ defined as the conjugate probability of the Beta distribution (prior) follows a Bernoulli distribution. With that in mind, the resulting user-file partition is reminiscent to that of the \ac{CRP} \cite{crp}. \ac{CRP} is based upon a metaphor in which the objects are customers in a restaurant, and the classes are the tables at which they sit. In particular, in a restaurant with a large number of tables, each with an infinite number of seats, customers enter the restaurant one after another, and each chooses a table at random. In the \ac{CRP} with parameter $\beta$, each customer chooses an occupied table with a probability proportional to the number of occupants, and chooses the next vacant table with probability proportional to $\beta$. Specifically, the first customer chooses the first table with probability $\frac{\beta}{\beta}=1$. The second customer chooses the first table with probability $\frac{1}{1+\beta}$, and the second table with probability $\frac{\beta}{1+\beta}$. After the second customer chooses the second table, the third customer chooses the first table with probability $\frac{1}{2+\beta}$, the second table with probability $\frac{1}{2+\beta}$ and the third table with probability $\frac{\beta}{2+\beta}$. This process continues until all customers have seats, defining a distribution over allocations of people to tables. Therefore, the decisions of subsequent customers are influenced by the previous customers' feedbacks, in which customers learn from the previous selections to update their beliefs on the files and the probabilities with which they choose their files. 

In view of this, the content dissemination in the social network is analogous to the table selection in a \ac{CRP}. In fact, if we view the overlay network as a Chinese restaurant, the contents as the very large number of files, and the users as the customers, we can interpret the contents dissemination process online by a \ac{CRP}. That is within every social community, users sequentially request to download their sought-after content, and when a user downloads its content, the recorded hits are recorded (i.e., history). In turn, this action affects the probability that this content will be requested by others users within the same social community, where popular contents are requested more frequently and new contents less frequently. Let $\bold{Z}_{N \times F}$ be a random binary matrix indicating which contents are selected by each user, where $z_{nf}=1$ if user $n$ selects content $f$ and $0$ otherwise. It can be shown that  \cite{crp}:
\begin{equation}
P(\bold{Z})=\frac{\beta^{F_h}\Gamma(\beta)}{\Gamma(\beta+N)}\prod_{f=1}^{F_h} (m_{f}-1)!
\end{equation}
in which $\Gamma(.)$ is the Gamma function, $m_f$ is the number of users currently assigned to content $f$ (or viewing history) and $F_h$ is the set of contents with viewing histories with $m_f>0$.
Therefore, for a given $P(\bold{Z})$, the popular files in each community can be stored greedily in the cache of influential users.
%%%%%%%%%%%%%%%%%%%%%%%%%%%%%%%%%% Case Study 2: Numerical results and discussion
\subsection{Numerical results and discussion}
The experimental setup is made of $N$ users connected to $M$ small cells. Each user is connected to its SCBS via a wireless link, and its neighbours via \ac{D2D} links. The total wireless link capacity of \glspl{SBS} is $C_w$ and the total \ac{D2D} link capacity is $C_d$. In order to see the impact of the parameters of interest, wireless link capacities are divided equally among users and the total \ac{D2D} link capacities are shared according to users' social links. The evaluation metrics are similar to those in case study I. The social-aware proactive caching is carried out as follows: If the requested file exists in neighbours' \ac{D2D} caches, the user is simultaneously served from the \ac{SBS} and its neighbours according to the available link capacities. A file request is said to be satisfied if the delivery time is below the threshold. The \emph{small cell load} is the amount of small cells' bandwidth consumed by the users over the total consumed bandwidth. All parameters are summarized in Table \ref{tab:setup_params2}.

At $t=0$, the arrival times of requests and their corresponding users are sampled uniformly at random for a time interval $T$. The social network is synthetically generated using the preferential attachment model \cite{NetworksNewman}. The $K$-most influential users are inferred using  eigenvector centrality, then, communities are formed via $K$-means clustering \cite{KMeans}. Subsequently, within every social community, the file popularity distribution is sampled from the $\textrm{CRP}(\beta)$ and proactive caching  is  carried out by storing popular files of the community. Random caching is used for comparison purposes.

\begin{figure*}[!ht]
\centering
%%%%%%%%%%%%%%%%%%%%%%%%%%%%%%%%%%%% Plots Case 2: Satisfied requests
\begin{tikzpicture}[scale=0.68, baseline]
	\begin{axis}[
		grid = major,
		cycle list name=thechairLivingOnTheEdge,
		mark repeat={4},
		ymin=0.65,ymax=1.05, ytick={0.7, 0.8, 0.9, 1},
		xmin=-0.05,xmax=1.05, xtick={0, 0.2, 0.4, 0.6, 0.8, 1},
	  	legend columns=4,
		legend entries={Proactive (Low load), Proactive (High load), Reactive (Low load), Reactive (High load)},
		legend cell align=left,
	  legend style ={font=\tiny},
		legend to name=namedmethods,
		ylabel = Satisfied requests]
		
	\addplot plot coordinates {
		(0.066667,	0.99997)
		(0.13333,	0.99961)
		(0.2,	0.99969)
		(0.26667,	0.99972)
		(0.33333,	0.99972)
		(0.4,	0.99944)
		(0.46667,	0.99944)
		(0.53333,	0.9995)
		(0.6,	0.99942)
		(0.66667,	0.99936)
		(0.73333,	0.99938)
		(0.8,	0.99929)
		(0.86667,	0.99921)
		(0.93333,	0.99917)
		(1,	0.99908)
	}; 	

	\addplot plot coordinates {
		(0.066667,	0.98769)
		(0.13333,	0.97658)
		(0.2,	0.96087)
		(0.26667,	0.94779)
		(0.33333,	0.93369)
		(0.4,	0.92141)
		(0.46667,	0.912)
		(0.53333,	0.89769)
		(0.6,	0.88634)
		(0.66667,	0.87426)
		(0.73333,	0.86167)
		(0.8,	0.84874)
		(0.86667,	0.83563)
		(0.93333,	0.82287)
		(1,	0.80753)
	}; 

	\addplot plot coordinates {
		(0.066667,	0.99244)
		(0.13333,	0.98223)
		(0.2,	0.97137)
		(0.26667,	0.96023)
		(0.33333,	0.94989)
		(0.4,	0.94122)
		(0.46667,	0.93201)
		(0.53333,	0.92189)
		(0.6,	0.91227)
		(0.66667,	0.9039)
		(0.73333,	0.89569)
		(0.8,	0.88754)
		(0.86667,	0.8771)
		(0.93333,	0.86946)
		(1,	0.85779)
	}; 

	\addplot plot coordinates {
		(0.066667,	0.98458)
		(0.13333,	0.96754)
		(0.2,	0.94932)
		(0.26667,	0.93055)
		(0.33333,	0.91324)
		(0.4,	0.89672)
		(0.46667,	0.88257)
		(0.53333,	0.86475)
		(0.6,	0.84942)
		(0.66667,	0.83331)
		(0.73333,	0.81639)
		(0.8,	0.79984)
		(0.86667,	0.78054)
		(0.93333,	0.76379)
		(1,	0.74415)
	}; 

	\end{axis}
\end{tikzpicture}
%%%%%%%%%%%%%%%%%%
\begin{tikzpicture}[scale=0.68, baseline]
	\begin{axis}[
		grid = major,
		cycle list name=thechairLivingOnTheEdge,
		ymin=0.65,ymax=1.05, ytick={0.7, 0.8, 0.9, 1},
		xmin=-0.05,xmax=1.05, xtick={0, 0.2, 0.4, 0.6, 0.8, 1},
		mark repeat={4}]
		
	\addplot plot coordinates {
		(0,	0.85214)
		(0.066667,	0.9645)
		(0.13333,	0.98563)
		(0.2,	0.9943)
		(0.26667,	0.99758)
		(0.33333,	0.999)
		(0.4,	0.99947)
		(0.46667,	0.99994)
		(0.53333,	0.99994)
		(0.6,	1)
		(0.66667,	1)
		(0.73333,	1)
		(0.8,	1)
		(0.86667,	1)
		(0.93333,	1)
		(1,	1)
	};

	\addplot plot coordinates {
		(0,	0.71591)
		(0.066667,	0.78462)
		(0.13333,	0.82931)
		(0.2,	0.86261)
		(0.26667,	0.88557)
		(0.33333,	0.90737)
		(0.4,	0.92439)
		(0.46667,	0.9377)
		(0.53333,	0.9492)
		(0.6,	0.96021)
		(0.66667,	0.969)
		(0.73333,	0.97741)
		(0.8,	0.9834)
		(0.86667,	0.99093)
		(0.93333,	0.99585)
		(1,	1)
	}; 

	\addplot plot coordinates {
		(0,	0.85214)
		(0.066667,	0.86376)
		(0.13333,	0.88875)
		(0.2,	0.89445)
		(0.26667,	0.89911)
		(0.33333,	0.91984)
		(0.4,	0.92721)
		(0.46667,	0.96556)
		(0.53333,	0.9693)
		(0.6,	0.97416)
		(0.66667,	0.98268)
		(0.73333,	0.98446)
		(0.8,	0.99349)
		(0.86667,	0.99499)
		(0.93333,	0.99658)
		(1,	1)
	}; 

	\addplot plot coordinates {
		(0,	0.71591)
		(0.066667,	0.73137)
		(0.13333,	0.76129)
		(0.2,	0.78212)
		(0.26667,	0.80163)
		(0.33333,	0.83186)
		(0.4,	0.85318)
		(0.46667,	0.88519)
		(0.53333,	0.90332)
		(0.6,	0.91995)
		(0.66667,	0.93545)
		(0.73333,	0.95033)
		(0.8,	0.96689)
		(0.86667,	0.97801)
		(0.93333,	0.98877)
		(1,	1)
	}; 

	\end{axis}
\end{tikzpicture}
%%%%%%%%%%%%%%%%%%
\begin{tikzpicture}[scale=0.68, baseline]
	\begin{axis}[
		grid = major,
		cycle list name=thechairLivingOnTheEdge,
		ymin=0.65,ymax=1.05, ytick={0.7, 0.8, 0.9, 1},
		xmin=-0.05,xmax=1.05, xtick={0, 0.2, 0.4, 0.6, 0.8, 1},
		mark repeat={4}]
		
	\addplot plot coordinates {
		(0,	1)
		(0.066667,	1)
		(0.13333,	0.99895)
		(0.2,	0.99453)
		(0.26667,	0.99061)
		(0.33333,	0.98454)
		(0.4,	0.98146)
		(0.46667,	0.97725)
		(0.53333,	0.97416)
		(0.6,	0.96977)
		(0.66667,	0.96841)
		(0.73333,	0.96564)
		(0.8,	0.96239)
		(0.86667,	0.96094)
		(0.93333,	0.95911)
		(1,	0.95609)
	}; 	

	\addplot plot coordinates {
		(0,	1)
		(0.066667,	0.98062)
		(0.13333,	0.94502)
		(0.2,	0.91213)
		(0.26667,	0.89117)
		(0.33333,	0.87843)
		(0.4,	0.86687)
		(0.46667,	0.85044)
		(0.53333,	0.83506)
		(0.6,	0.82723)
		(0.66667,	0.82088)
		(0.73333,	0.81396)
		(0.8,	0.81258)
		(0.86667,	0.80967)
		(0.93333,	0.8094)
		(1,	0.80375)
	}; 

	\addplot plot coordinates {
		(0,	0.92717)
		(0.066667,	0.92551)
		(0.13333,	0.92476)
		(0.2,	0.92637)
		(0.26667,	0.92851)
		(0.33333,	0.92931)
		(0.4,	0.9267)
		(0.46667,	0.92456)
		(0.53333,	0.92742)
		(0.6,	0.92888)
		(0.66667,	0.92905)
		(0.73333,	0.92344)
		(0.8,	0.92386)
		(0.86667,	0.92429)
		(0.93333,	0.92359)
		(1,	0.92214)
	}; 

	\addplot plot coordinates {
		(0,	0.77563)
		(0.066667,	0.75128)
		(0.13333,	0.75953)
		(0.2,	0.75483)
		(0.26667,	0.75481)
		(0.33333,	0.74977)
		(0.4,	0.74542)
		(0.46667,	0.75336)
		(0.53333,	0.75234)
		(0.6,	0.74992)
		(0.66667,	0.75165)
		(0.73333,	0.74926)
		(0.8,	0.75026)
		(0.86667,	0.74955)
		(0.93333,	0.74742)
		(1,	0.74581)
	}; 

	\end{axis}
\end{tikzpicture}
%%%%%%%%%%%%%%%%%%%%%%%%%%%%%%%%%%%% Plots Case 2: Small cell load
\begin{tikzpicture}[scale=0.68, baseline]
	\begin{axis}[
		grid = major,
		cycle list name=thechairLivingOnTheEdge,
		mark repeat={4},
		ymin=0.35,ymax=1.05, ytick={0.4, 0.6, 0.8, 1},
		xmin=-0.05,xmax=1.05, xtick={0, 0.2, 0.4, 0.6, 0.8, 1},		
		xlabel=Number of requests ($\widehat{R}$),
		ylabel=Small cell load]
		
	\addplot plot coordinates {
		(0.066667,	0.4062)
		(0.13333,	0.41006)
		(0.2,	0.40736)
		(0.26667,	0.40552)
		(0.33333,	0.40265)
		(0.4,	0.40043)
		(0.46667,	0.39719)
		(0.53333,	0.39411)
		(0.6,	0.39192)
		(0.66667,	0.38952)
		(0.73333,	0.38754)
		(0.8,	0.38562)
		(0.86667,	0.38343)
		(0.93333,	0.38136)
		(1,	0.37923)
	};	

	\addplot plot coordinates {
		(0.066667,	0.78089)
		(0.13333,	0.79967)
		(0.2,	0.79918)
		(0.26667,	0.79474)
		(0.33333,	0.79305)
		(0.4,	0.79105)
		(0.46667,	0.78814)
		(0.53333,	0.78754)
		(0.6,	0.78469)
		(0.66667,	0.78327)
		(0.73333,	0.78288)
		(0.8,	0.78421)
		(0.86667,	0.78291)
		(0.93333,	0.78258)
		(1,	0.78195)
	};

	\addplot plot coordinates {
		(0.066667,	0.69473)
		(0.13333,	0.68665)
		(0.2,	0.68396)
		(0.26667,	0.68376)
		(0.33333,	0.67952)
		(0.4,	0.68092)
		(0.46667,	0.68151)
		(0.53333,	0.68001)
		(0.6,	0.6807)
		(0.66667,	0.6794)
		(0.73333,	0.67968)
		(0.8,	0.6779)
		(0.86667,	0.67765)
		(0.93333,	0.67762)
		(1,	0.6768)
	};

	\addplot plot coordinates {
		(0.066667,	0.9189)
		(0.13333,	0.91642)
		(0.2,	0.91611)
		(0.26667,	0.91435)
		(0.33333,	0.90975)
		(0.4,	0.90701)
		(0.46667,	0.90464)
		(0.53333,	0.90387)
		(0.6,	0.90175)
		(0.66667,	0.90096)
		(0.73333,	0.90162)
		(0.8,	0.90172)
		(0.86667,	0.9011)
		(0.93333,	0.90013)
		(1,	0.90038)
	};

	\end{axis}
\end{tikzpicture}
%%%%%%%%%%%%%%%%%%
\begin{tikzpicture}[scale=0.68, baseline]
	\begin{axis}[
		grid = major,
		cycle list name=thechairLivingOnTheEdge,
		mark repeat={4},
		ymin=0.35,ymax=1.05, ytick={0.4, 0.6, 0.8, 1},
		xmin=-0.05,xmax=1.05, xtick={0, 0.2, 0.4, 0.6, 0.8, 1},		
		xlabel=Cache size ($\widehat{S}$)]
		
	\addplot plot coordinates {
		(0,	1)
		(0.066667,	0.55183)
		(0.13333,	0.46037)
		(0.2,	0.41742)
		(0.26667,	0.40479)
		(0.33333,	0.39831)
		(0.4,	0.3959)
		(0.46667,	0.39484)
		(0.53333,	0.39435)
		(0.6,	0.39402)
		(0.66667,	0.39402)
		(0.73333,	0.39402)
		(0.8,	0.39402)
		(0.86667,	0.39402)
		(0.93333,	0.39402)
		(1,	0.39402)
	};	

	\addplot plot coordinates {
		(0,	1)
		(0.066667,	0.83065)
		(0.13333,	0.74971)
		(0.2,	0.68429)
		(0.26667,	0.6351)
		(0.33333,	0.58713)
		(0.4,	0.55272)
		(0.46667,	0.51894)
		(0.53333,	0.49412)
		(0.6,	0.46952)
		(0.66667,	0.45028)
		(0.73333,	0.42997)
		(0.8,	0.41539)
		(0.86667,	0.39803)
		(0.93333,	0.38743)
		(1,	0.37755)
	};

	\addplot plot coordinates {
		(0,	1)
		(0.066667,	0.94695)
		(0.13333,	0.8528)
		(0.2,	0.8253)
		(0.26667,	0.8018)
		(0.33333,	0.70901)
		(0.4,	0.68098)
		(0.46667,	0.54864)
		(0.53333,	0.53141)
		(0.6,	0.51166)
		(0.66667,	0.47393)
		(0.73333,	0.46898)
		(0.8,	0.4289)
		(0.86667,	0.4204)
		(0.93333,	0.41123)
		(1,	0.39402)
	};

	\addplot plot coordinates {
		(0,	1)
		(0.066667,	0.93971)
		(0.13333,	0.87197)
		(0.2,	0.83177)
		(0.26667,	0.79586)
		(0.33333,	0.74208)
		(0.4,	0.70011)
		(0.46667,	0.63681)
		(0.53333,	0.60074)
		(0.6,	0.5658)
		(0.66667,	0.5286)
		(0.73333,	0.49798)
		(0.8,	0.46203)
		(0.86667,	0.43646)
		(0.93333,	0.4105)
		(1,	0.37755)
	};

	\end{axis}
\end{tikzpicture}
%%%%%%%%%%%%%%%%%%
\begin{tikzpicture}[scale=0.68, baseline]
	\begin{axis}[
		grid = major,
		cycle list name=thechairLivingOnTheEdge,
		mark repeat={4},
		ymin=0.35,ymax=1.05, ytick={0.4, 0.6, 0.8, 1},
		xmin=-0.05,xmax=1.05, xtick={0, 0.2, 0.4, 0.6, 0.8, 1},		
		xlabel=CRP parameter ($\widehat{\beta}$)]
		
	\addplot plot coordinates {
		(0,	0.39402)
		(0.066667,	0.39413)
		(0.13333,	0.399)
		(0.2,	0.41579)
		(0.26667,	0.4365)
		(0.33333,	0.46447)
		(0.4,	0.47724)
		(0.46667,	0.49692)
		(0.53333,	0.51015)
		(0.6,	0.52913)
		(0.66667,	0.53465)
		(0.73333,	0.5475)
		(0.8,	0.56264)
		(0.86667,	0.56671)
		(0.93333,	0.57084)
		(1,	0.58429)
	};	

	\addplot plot coordinates {
		(0,	0.37755)
		(0.066667,	0.42342)
		(0.13333,	0.50299)
		(0.2,	0.57803)
		(0.26667,	0.62413)
		(0.33333,	0.65185)
		(0.4,	0.6759)
		(0.46667,	0.70946)
		(0.53333,	0.73852)
		(0.6,	0.75324)
		(0.66667,	0.76556)
		(0.73333,	0.77816)
		(0.8,	0.78197)
		(0.86667,	0.78724)
		(0.93333,	0.78785)
		(1,	0.79795)
	};

	\addplot plot coordinates {
		(0,	0.66931)
		(0.066667,	0.69374)
		(0.13333,	0.68569)
		(0.2,	0.69374)
		(0.26667,	0.69335)
		(0.33333,	0.69776)
		(0.4,	0.70254)
		(0.46667,	0.70637)
		(0.53333,	0.6997)
		(0.6,	0.70216)
		(0.66667,	0.70205)
		(0.73333,	0.71399)
		(0.8,	0.71524)
		(0.86667,	0.70587)
		(0.93333,	0.71012)
		(1,	0.7154)
	};

	\addplot plot coordinates {
		(0,	0.85398)
		(0.066667,	0.89632)
		(0.13333,	0.87323)
		(0.2,	0.88294)
		(0.26667,	0.8865)
		(0.33333,	0.88911)
		(0.4,	0.89871)
		(0.46667,	0.89237)
		(0.53333,	0.88828)
		(0.6,	0.8946)
		(0.66667,	0.89393)
		(0.73333,	0.89879)
		(0.8,	0.90131)
		(0.86667,	0.89908)
		(0.93333,	0.90307)
		(1,	0.90317)
	};

	\end{axis}
\end{tikzpicture}
%%%%%%%%%%%%%%%%%%%%%%%%%%%%%%%%%%%% Parameters
\begin{tikzpicture}[scale=0.68, baseline]
\node (table) {
\scriptsize
\resizebox{0.29\linewidth}{!}{
\begin{tabular}{l*{3}{c}}
& Low load & High load \\
\hline
${\widehat S}$ 	& $0.4$ & $0.1$ \\
${\widehat \beta}$  & $0.1$ & $0.9$ \\
\hline
\end{tabular}
}
};
\end{tikzpicture}
%%%%%%%%%%%%%%%%%%
\begin{tikzpicture}[scale=0.68, baseline]
\node (table) {
\resizebox{0.29\linewidth}{!}{
\scriptsize
\begin{tabular}{l*{3}{c}}
& Low load & High load \\
\hline
${\widehat R}$ 			& $0.5$ & $0.98 $ \\
${\widehat \beta}$  & $0.1$ & $0.9$   \\
\hline
\end{tabular}
}
};
\end{tikzpicture}
%%%%%%%%%%%%%%%%%%
\begin{tikzpicture}[scale=0.68, baseline]
\node (table) {
\resizebox{0.29\linewidth}{!}{
\scriptsize
\begin{tabular}{l*{3}{c}}
& Low load &  High load \\
\hline
${\widehat R}$ 			& $0.5$ & $0.98 $ \\
${\widehat S}$ 			& $0.4$ & $0.1$ \\
\hline
\end{tabular}
}
};
\end{tikzpicture}
\\
\ref{namedmethods}
\caption{Social-Aware Caching via D2D: Evolutions of satisfied requests  and small cell load with respect to number of requests ${\widehat R}$, cache size ${\widehat S}$ and CRP concentration parameter ${\widehat \beta}$.}
\label{fig:sim-case2}
\end{figure*}
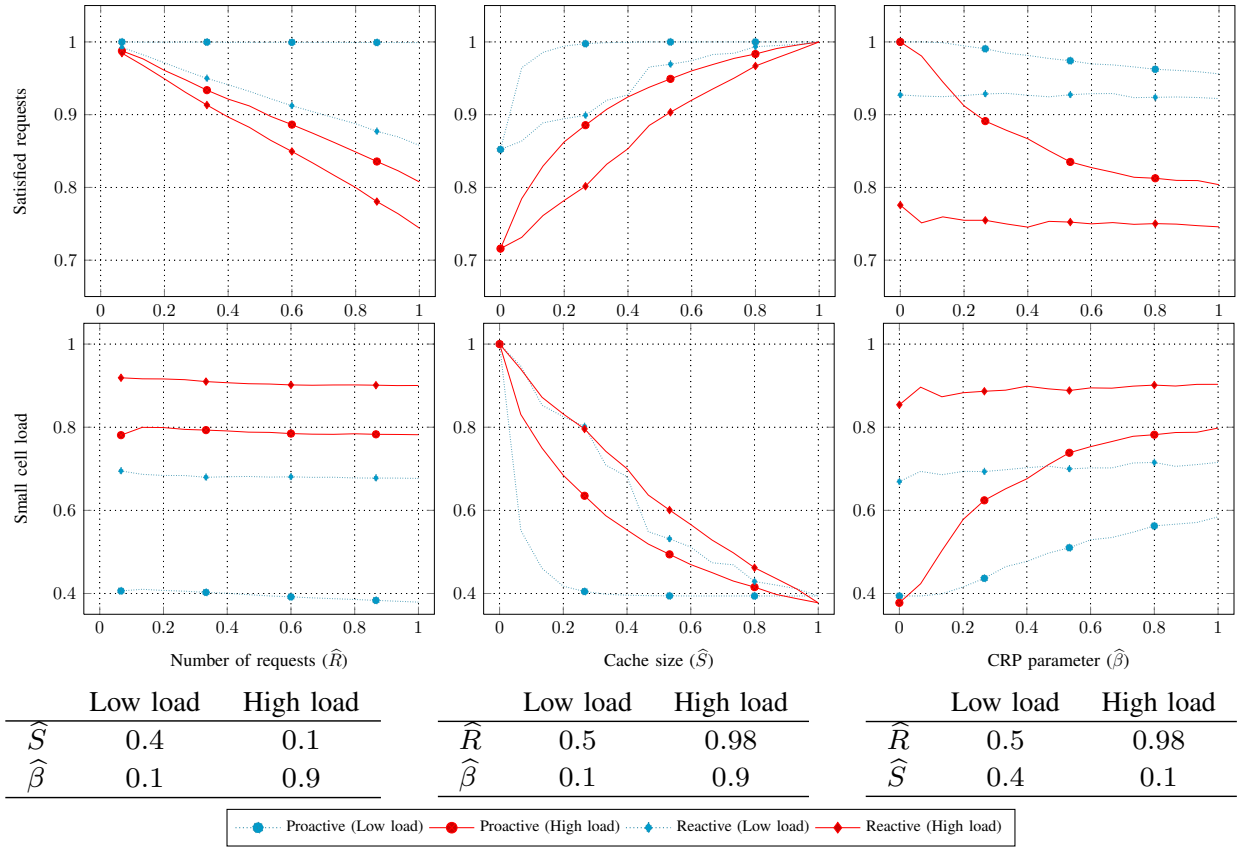

Three parameters are of interest: (i) number of requests $R$, \ac{D2D} cache size $S$ and \ac{CRP}  parameter $\beta$. These parameters are normalized by ${R^{\star}}$, $L \times F$, and $100$ respectively, and shown as ${\widehat R}$, ${\widehat S}$ and ${\widehat \beta}$. The performance evaluation of satisfied requests and backhaul load with respect to these parameters is plotted in Fig. \ref{fig:sim-case2}. 
\begin{table}[ht]
\centering
\begin{tabular}{|c|l|c|}
\hline
\textbf{Parameter}			  &  \textbf{Description} 	& 	\textbf{Value}\\
\hline
$T$    		& Time slots							& 	$ 1024$ seconds \\
\hline
$M$   		& Number of small cells 				& 	$4$ \\
\hline
$K$ 			& Number of communities 				& 	$3$ 		\\
\hline
$N$ 			& Number of user terminals 			& 	$32$ 	\\
\hline
$F$			& Number of files 					&  $128$		\\
\hline
$L$	 		& Length of each file 				& 	$1$ Mbit	\\
\hline
$B$ 			& Bitrate of each  file				&	$1$ Mbit/s	\\
\hline
$C_w$ 		& Total \glspl{SBS} link capacity 		&	$32$ Mbit/s	 \\
\hline
$C_b$ 		& Total \ac{D2D} link capacity		&	$64$	 Mbit/s	\\
\hline
\hline
$R^{\star}$	& 	Maximum number of requests 		&	$9464$ 	\\
\hline
\hline
$R$ 			& 	Number of requests 				&	$0 \sim 9464$		\\
\hline
$S$    		& Total \ac{D2D} cache size			& 	$0 \sim 128$ MBit	\\
\hline
$\beta$   	& \ac{CRP} parameter	 				& 	$0 \sim 100$			\\
\hline
\end{tabular}
\vspace{3mm}
\caption{List of parameters for case study II.}
\label{tab:setup_params2}
\end{table}
As ${\widehat R}$ increases, the number of satisfied requests increases rapidly while the small cell load decreases in a very low pace. The proactive caching approach outperforms the reactive approach in all load regimes. On the other hand, as ${\widehat S}$ increases, the gains of the satisfaction increases and backhaul load decreases, non-linearly.

In the case of an increment of $\beta$, which means that the number of distinct files is growing, the satisfaction and the backhaul load are approximately becoming constant in the reactive approach. The proactive approach has a better performance, but it gets closer to the reactive one as  $\beta$ grows. As mentioned previously, this is because of the growing catalog size where the cache size is fixed.
%
%%%%%%%%%%%%%%%%%%%%%%%%%%%%%%%%%%%%%%%%%%%%%%%%%%%%%%%%%%%%%%%%%%%
%%%%%%%%%%%%%%%%%%%%%%%%%%%%%%%%%% CONCLUSION
\section{Conclusion}
\label{sec:conclusions}
In this article, we discussed the limitations of current reactive networks and proposed a novel proactive networking paradigm where caching plays a crucial role. By exploiting the predictive capabilities of $5$G networks, coupled with notions of context-awareness and social networks, it was shown that peak data traffic demands can be substantially reduced by proactively serving predictable users demands, via caching strategic contents at both the base station and user's devices. This predictive networking, with adequate storage capabilities at the edge of the network, holds the promise of helping mobile operators tame the data tsunami, which will continue straining current networks.

The proactive caching paradigm, which is still in its infancy, has been mainly investigated from an upper layer perspective. An interesting future work would be exploiting multicast gains and designing intelligent coding schemes which take into account cross-layer issues. Yet another line of investigation is the joint optimization of proactive content caching, interference management and scheduling techniques.  In terms of resource allocation, what contents to store where, given heterogeneous content popularity, how to match users’ requests to base stations with optimal replication ratios are of high interest for optimal heterogeneous load balancing. In cases of mobility, smarter mechanisms are required in which \glspl{SBS} need to coordinate to do a joint load balancing and content sharing. Lastly, one can formulate the proactive caching problem from a game theoretic learning perspective where \ac{SBS} minimize the cache miss by striking a good balance between cached contents that will be requested and contents not cached but requested by users. This is also referred to as \emph{exploration vs. exploitation} paradigm.
%%%%%%%%%%%%%%%%%%%%%%%%%%%%%%%%%%%%%%%%%%%%%%%%%%%%%%%%%%%%%%%%%%%
%%%%%%%%%%%%%%%%%%%%%%%%%%%%%%%%%% BIBLIOGRAPHY
\bibliographystyle{IEEEtran}
\bibliography{references}
\end{document}